\documentclass[aps,prb,twocolumn,reprint,superscriptaddress,showkeys,floatfix,amsmath,amssymb]{revtex4-2}
\usepackage[utf8]{inputenc}
\usepackage{placeins}
\usepackage{amsmath}
\usepackage[utf8]{inputenc} 
\DeclareUnicodeCharacter{202F}{\,} 
\usepackage{newunicodechar}
\newunicodechar{ }{\,}
\newunicodechar{∼}{\sim}
\usepackage{color}
\usepackage{graphicx}
\usepackage{xcolor, soul}
\usepackage{dcolumn}
\usepackage{natbib}
\usepackage{natbib}
\usepackage{bm}
\usepackage[update,prepend]{epstopdf}
\usepackage[caption=false]{subfig} 
\usepackage{amsmath}
\usepackage{natbib}
\usepackage{physics}
\usepackage{xr}
\usepackage{amsmath}
\usepackage{dcolumn}
\usepackage{booktabs}
\usepackage[normalem]{ulem}
\usepackage{hyperref}
\begin{document}
	
	\preprint{APS/123-QED}
	
	\title{Electronic properties of Kagome metal YbV$_3$Sb$_4$: A First-Principles Study }
	
	
	\author{D. Gurung}
	\affiliation{Department of Physics, Mizoram University, Aizawl-796004, India} 
	
	\author{Keshav Shrestha}
	\affiliation{Department of Chemistry and Physics, West Texas A\&M University, Canyon, Texas 79016, USA }
	\author{Shalika R. Bhandari} 
	\affiliation{Department of Physics, Bhairahawa Multiple Campus,  Tribhuvan University, Siddarthanagar-32900, Rupandehi, Nepal}
	
	\author{Samy Brahimi}
	\affiliation{Laboratoire de Physique et Chimie Quantique, Universite Mouloud Mammeri de Tizi-Ouzou, 15000 Tizi-Ouzou, Algeria}
	\affiliation{Peter Gr\"unberg Institute, Forschungszentrum J\"ulich and JARA, J\"ulich, Germany}
	\author{Samir Lounis}
	\affiliation{Institute of Physics, Martin-Luther-University Halle-Wittenberg, 06099 Halle (Saale), Germany}
	\author{D. P. Rai}
	\email[D. P. Rai:]{ dibyaprakashrai@gmail.com}
	\affiliation{Department of Physics, Mizoram University, Aizawl-796004, India}
	\affiliation{Peter Gr\"unberg Institute, Forschungszentrum J\"ulich and JARA, J\"ulich, Germany}
	\date{\today}
	
	\begin{abstract}
		We have investigated the vanadium-based Kagome metal YbV$_3$Sb$_4$ using density functional theory (DFT) combined with the Wannier function analysis. We explore the electronic properties, de Haas–van Alphen (dHvA) effect and Fermi surface. Our calculations reveal the metallic characteristic in which majority of the states around the Fermi energy (E$_F$) is contributed by the V-$3d$ orbitals, while the localized Yb-$4f$ states positioned below it. The inclusion of spin-orbit coupling SOC induces the splitting of Yb-4$f$ states, while its impact on the V-3$d$ states is moderate. 
		Furthermore, we have incorporated 
		SOC+U, where U being the Hubbard parameter, which drastically changes the Yb-4$f$ states creating additional splitting leading to three distinct peaks in the density of states (DOS). Meanwhile, the V-3$d$ atoms with the Kagome lattice contribute maximum to the transport properties, exhibits flat bands near the E$_F$ while being protected under SOC and U+SOC. Herein, we report the vulnerability of the Yb-4$f$ states under SOC and U+SOC. 
		Furthurmore, The Fermi surface is found to comprise of quasi-2D cylindrical sheets centered at the $\Gamma$ point, along with smaller pockets near the Brillouin zone boundaries, which under combined U+SOC, a small spherical pocket emerges and the cylindrical sheet exhibits slight deformations. The  dHvA frequencies reach as high as 70 kilotesla, which increase  with tilt angle, exhibiting a nearly parabolic trend  as expected for cylindrical orbits, while a low-frequency branch remains below 1 kT.Only the U+SOC case shows noticeable modification in both the Fermi surface and the dHvA oscillation.  
		Crucially, the $Z_2$ invariant calculation identifies YbV$_3$Sb$_4$ as a strong topological metal ($\nu_0 = 1$).
		These findings not only advance our understanding of the underlying quantum phenomena in rare-earth Kagome systems, but also establish YbV$_3$Sb$_4$ as a compelling and promising platform for exploring intertwined topology and electron correlations in kagome lattices, thereby offering valuable insights for engineering quantum phases in layered materials.\\

	\end{abstract}
	
	\maketitle
	
	
	\section{INTRODUCTION}
	
	Kagome materials have been of intensive research interest recently in solid-state physics due to their unique electronic properties, as they provide an opening into various quantum phenomena such as charge density waves \cite{teng}, non-trivial topology \cite{Jiangtop}, and superconductivity among a few \cite{jiang23,neupert2022charge}. The lattice geometry of Kagome materials comprising the Japanese basket weave pattern of interlaced triangles and corner sharing triangles with geometrically frustrated two-dimensional (2D) lattices is responsible for the creation of the flat band in the electronic structure, van Hove singularities, and Dirac cones naturally \cite{hastings,mazin,volkov,wang,wang2025higher}.It also displays tunable Dirac band gaps \cite{ye2018,yin2018} , Chern number, Chern gaps \cite{yin20} 
	It was recently noted that acoustic Kagome lattices exhibit higher-order non-Hermitian skin effects (NHSE)\cite{zhong2025higher}, and in topological invariants such as the $Z_{2}$ invariant \cite{kondo2019z,joshi2019z,bhandari2024first}, which further illustrate the interaction between geometry and topology in these systems.
	The Fermi surface reconstructions revealed by de Haas-van Alphen (dHvA) studies have been used to investigate electronic correlations of Kagome materials \cite{shtefiienko2025electronic,shtefiienko2025electronic2,dhital2024fermi,miertschin2025dhva,phillips2024fermi}, while nematicity and symmetry-breaking phenomena \cite{patra2025high,nie2022charge} emphasis the role of electronic interactions.
	
	Emergent quantum behaviors such as the Nernst effect \cite{asaba2021colossal,chen2022large,li2023enhanced} and the anomalous Hall effect \cite{ohgushi,liu2018giant,chen2021large,yu2021concurrence,wang2024orbital} are further revealed by transport measurements, signs of Berry curvature and chiral quasi-particles, and also unconventional superconductivity \cite{guguchia2023tunable,holbaek2023unconventional} that emerges under doping or pressure. These findings position Kagome materials at the frontier of quantum material research, bridging topology, frustration, and strong correlations.
	
	In addition to the research on the frustrated quantum magnetism of Kagome materials, the recent research interest has been directed towards their electronic properties, magnetic and topological quantum transport properties. Several materials have been discovered to host Kagome lattices and novel Kagome physics \cite{jovanovic}. Some of the examples having Kagome-type lattices are Fe$_{3}$Sn$_{2}$ which shows giant spin-orbit coupling \cite{ye2018,yin2018}, magnetic Weyl semimetals Co$_{3}$Sn$_{2}$S$_{2}$ \cite{liu2019,liu2018,morali2019}, charge density wave (CDW) ordering compound ScV$_6$Sn$_6$ \cite{arachchige2022charge}, Nematicity and Multigap Superconductivity in Titanium-Based Kagome Metal, CsTi$_3$Bi$_5$ \cite{yang2024superconductivity}, FeSn antiferromagnet \cite{kang2019},a skyrmion like triple-Q state in breathing Kagome lattices \cite{zhou2025triple}, Chern-gapped, Dirac fermions in ferromagnetic TbMn$_{6}$Sn$_{6}$ \cite{yin20}, the non-collinear antiferromagnet in Mn$_{3}$Sn \cite{kuroda2017},and substrate induced Kagome ordering in transition metal monolayers on h-BN \cite{zhou2024kagomerization}. Recently, a superconducting behaviour in the quasi-two-dimensional(2D) Kagome metals AV$_{3}$Sb$_{5}$ (A=K,Rb,and Cs)\cite{yinrb2021,ortiz2019,ortiz2020,ortiz2021,wilson2024v3sb5,cai2024angle} and high-temperature superconductivity in hydrogen-based systems AH$_3$Li$_5$ (A = C, Si, P) has been reported up to 120 K in PH$_3$Li$_5$ under compression \cite{liu2024proposed}. 
	
	A new type of vanadium-based Kagome bilayer, viz AV$_{6}$Sb$_{6}$, AV$_{3}$Sb$_{5}$ (A=K, Rb, Cs) and V$_{6}$Sb$_{4}$ with a generic chemical formula (A$_{m-1}$Sb$_{2m}$)(V$_{3}$Sb)$_{n}$ ($m=1,2$; $n=1,2$) has been reported \cite{shi2022} .
	Among these new compounds, AM$_{3}$X$_{4}$ consisting of (A: La, Ce, Sm, Ca, Yb, Eu, Gd;  M: V, Ti; X: Sb, Bi) are particularly fascinating. It is distinguished by the presence of zigzag chains made of A-site ions that run parallel to their somewhat distorted M-based Kagome sublattices~\cite{guo2023magnetic,ortiz2023ybv,chen2024tunable,motoyama2018magnetic,ovchinnikov2019bismuth,ortiz2023evolution}. Due to the wide variety of A cations available, this family includes both magnetic and non-magnetic compounds. Its crystal structure is significantly more layered, exfoliable, and quasi-2D, making it an intriguing subject for research.
	
	In this work, we study the compound YbV$_{3}$Sb$_{4}$, a recent addition to the larger AM$_{3}$X$_{4}$ family. Despite its intriguing structural motifs and the potential emergence of rare earth physics, YbV$_{3}$Sb$_{4}$ still remains mostly unexplored, with limited  experimental reports and first-principles studies to date. Its distinct structural motifs of vanadium-based Kagome nets scattered with zigzag chains of divalent Yb$^{2+}$ ions, make this compound containing rare-earth atoms an ideal candidate for a thorough investigation.
	Between 300 K and 60 mK, this compound is found to exhibit no discernible thermodynamic changes and is nonmagnetic~\cite{ortiz2023ybv}. Here, we present a first-principles study of the electronic characteristics of YbV$_{3}$Sb$_{4}$. 
	Based on density functional theory, we explore the dHvA effect, examine the Fermi surface, and utilize Wannier centers to compute the topological invariant $Z_2$.

	\section{COMPUTATIONAL DETAILS}
\begin{figure}[!ht]
		\centering
		\includegraphics[width=0.5\textwidth]{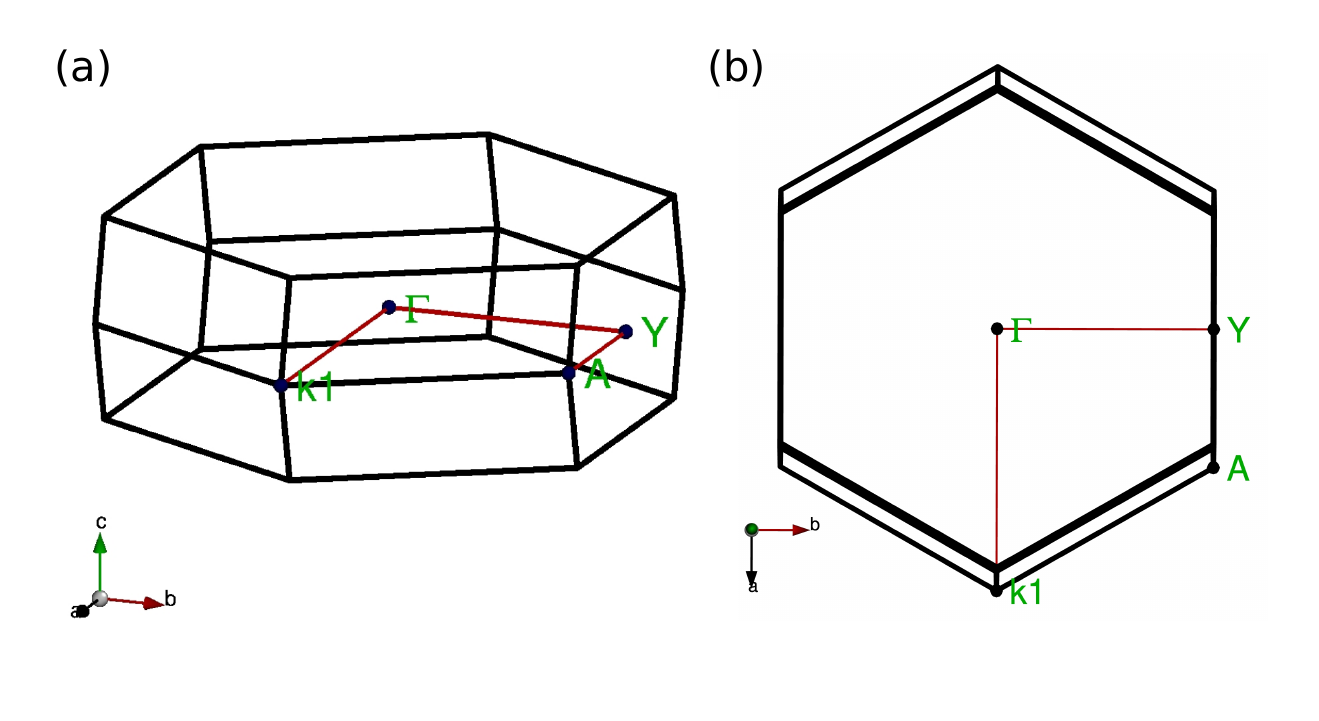} 
		\caption{(a)The path connecting the high symmetry points of the first Brillouin zone. (b)The top view of the Brillouin zone.}
		\label{fig:symmetry_points}
	\end{figure}

	Our ab initio calculations were performed using the full potential local orbital (FPLO) code version 22.00-62 \cite{koepernik1999full,fploweb}. The generalized gradient approximation (GGA) within the parameterization of Perdew, Burke and Ernzerhof was used for the exchange-correlation potential \cite{perdew1996generalized,PhysRevLett.78.1396}. To accurately account for the strong electron interactions among the strongly correlated V 3$d$ and Yb 4$f$ valence electrons in our system, effective Hubbard interaction ($U_{\text{eff}}$) within the DFT+U approximation has been applied, with fully screened Coulomb and exchange interaction parameters ($J$) \cite{vanadium1,vanadium2,ybvalue1,ybvalue2} as U$_{Yb}$=6 eV, J$_{Yb}$= 0.5374 eV, U$_V$=3.4000 eV and J$_V$=1.000 eV. Atomic positions were optimized until the residual forces on each atom fell below \(10^{-3}\,\mathrm{eV/\mathring{A}}\). The convergence criterion was set to \(10^{-8}\) for the charge density and \(10^{-8}\) Hartree for the total energy in all calculations. The integration over the Brillouin zone was performed using the Blöchl corrected linear tetrahedron method with a k-points grid of 12$\times$12$\times$12 points. This k-points mesh was used consistently across both scalar and fully relativistic calculations and for the evaluation of band structures and densities of states (DOS).  Full-relativistic effects including spin–orbit coupling (SOC) were treated in a full four-component Dirac formalism as implemented in FPLO \cite{koepernik1999full,fploweb}. The magnetization direction was fixed via the global spin quantization axis setting. Following the established methodology described in \cite{setyawan2010high,ortiz2023ybv} the high-symmetry k-path in the Brillouin zone was chosen. The high-symmetry $k$-path in the Brillouin zone, along which the band structures were computed, is shown in Fig.~\ref{fig:symmetry_points} and Table \ref{tab:table1}.
	
	\begin{table}[h!]
		\centering
		\caption{Coordinates of the high-symmetry points in the Brillouin zone (BZ) used for band structure calculations, expressed in units of $2\pi/a$, $2\pi/b$, and $2\pi/c$. }
		\begin{tabular}{|c|c|c|c|}
			\hline
			\textbf{Point} & \textbf{$k_x$ [$2\pi/a$]} & \textbf{$k_y$ [$2\pi/b$]} & \textbf{$k_z$ [$2\pi/c$]} \\ \hline
			$\Gamma$       & 0                         & 0                         & 0                         \\ \hline
			k1             & 0.69153348986391           & 0                         & 0                         \\ \hline
			A              & 0.3649862854056            & 0.57144308943088          & 0                         \\ \hline
			Y              & 0                         & 0.5714430894309           & 0                         \\ \hline
			$\Gamma$       & 0                         & 0                         & 0                         \\ \hline
		\end{tabular}
		h
		\label{tab:table1}
	\end{table}
	
	\begin{figure*}[htbp]
		\centering
		\includegraphics[width=\textwidth]{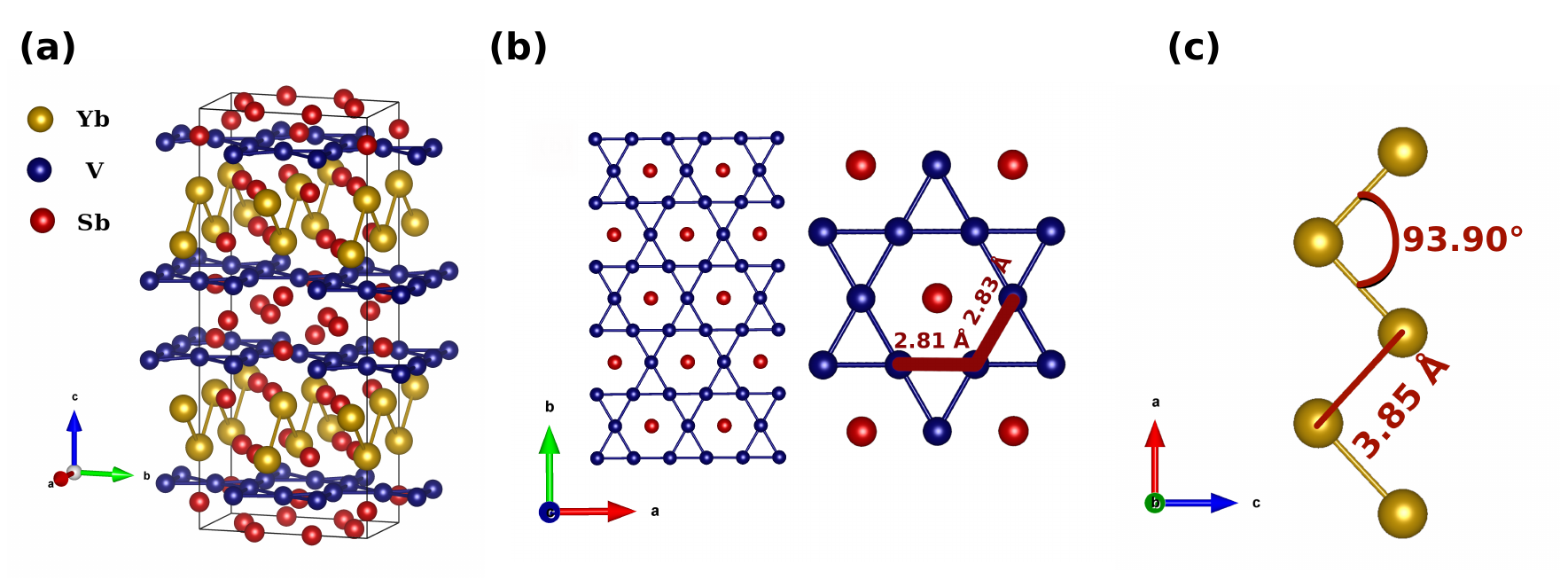}
		\caption{Crystal Structure of YbV$_3$Sb$_4$: (a) The unit cell of YbV$_3$Sb$_4$ includes Yb Kagome layers (shown in gold), with V atoms represented as blue spheres and Sb atoms as red spheres. (b) Crystal Structure of YbV$_3$Sb$_4$, highlighting the V-Based Kagome Network. (c) Structure of the Yb ions' zigzag sublattices.} 
		\label{fig:four_images}
	\end{figure*}
	
	The tight-binding Hamiltonian model was constructed using maximally projected Wannier functions, obtained through the PYFPLO module of the FPLO code~\cite{fploweb} [see FIG. S1 in Supplementary material]. The local Wannier basis set consisted of orbitals of Yb [$4f$, $5d$], V [$3d$, $4s$], and Sb [$5p$, $5s$]. To ensure consistency, the same 12$\times$12$\times$12 k-points mesh used in the self-consistent calculations was also applied to the Wannier model, in an energy window from $-6$ eV to $+5$ eV, including all states near the Fermi level. The Hamiltonian data were used in the $Z_2$ invariant calculations.
	
	\section{RESULTS AND DISCUSSION}
	\subsection{Structural Properties}
	
	YbV$_{3}$Sb$_{4}$ crystallizes in a face-centered orthorhombic lattice and belongs to the space group Fmmm (SG 69, Hall notation: $-\mathbf{F}22$)[see Fig.\ref{fig:four_images}]. The experimentally obtained~\cite{ortiz2023ybv} lattice parameters chosen for the calculation are $a = 5.623$ {\AA}, $b = 9.840$ {\AA}, $c = 23.652$ {\AA}. The corresponding point group is $D_{2h}$, indicating the presence of an inversion symmetry. The structure is symmorphic. The atomic positions are detailed in Table \ref{tab:atomic_positions}, following the corresponding Wyckoff positions.
	
	\begin{table}[ht!]
		\centering
		\caption{Atomic positions and symmetry labels for YbV$_{3}$Sb$_{4}$}
		\resizebox{\columnwidth}{!}{%
			\begin{tabular}{|c|c|c|c|c|c|c|}
				\hline
				\textbf{Element} & \textbf{Type} & \textbf{X} & \textbf{Y} & \textbf{Z} & \textbf{Wname} & \textbf{Symmetry} \\ \hline
				Yb & 70 & 0.00000  & -0.50000  & 0.30552  & 8i & mm2  \\ \hline
				Sb & 51 & 0.00000  & -0.50000  & 0.43536  & 8i & mm2  \\ \hline
				Sb & 51 & -0.50000 & -0.32904  & -0.50000 & 8h & m2m  \\ \hline
				Sb & 51 & -0.50000 & 0.34123   & 0.31426  & 16m & m..  \\ \hline
				V  & 23 & -0.50000 & -0.50000  & 0.40981  & 8i & mm2  \\ \hline
				V  & 23 & -0.25000 & 0.25000   & 0.40577  & 16j & ..2  \\ \hline
			\end{tabular}%
		}
		
		\label{tab:atomic_positions}
	\end{table}
	
	\begin{figure*}[h!]
		\centering
		\includegraphics[width=\textwidth]{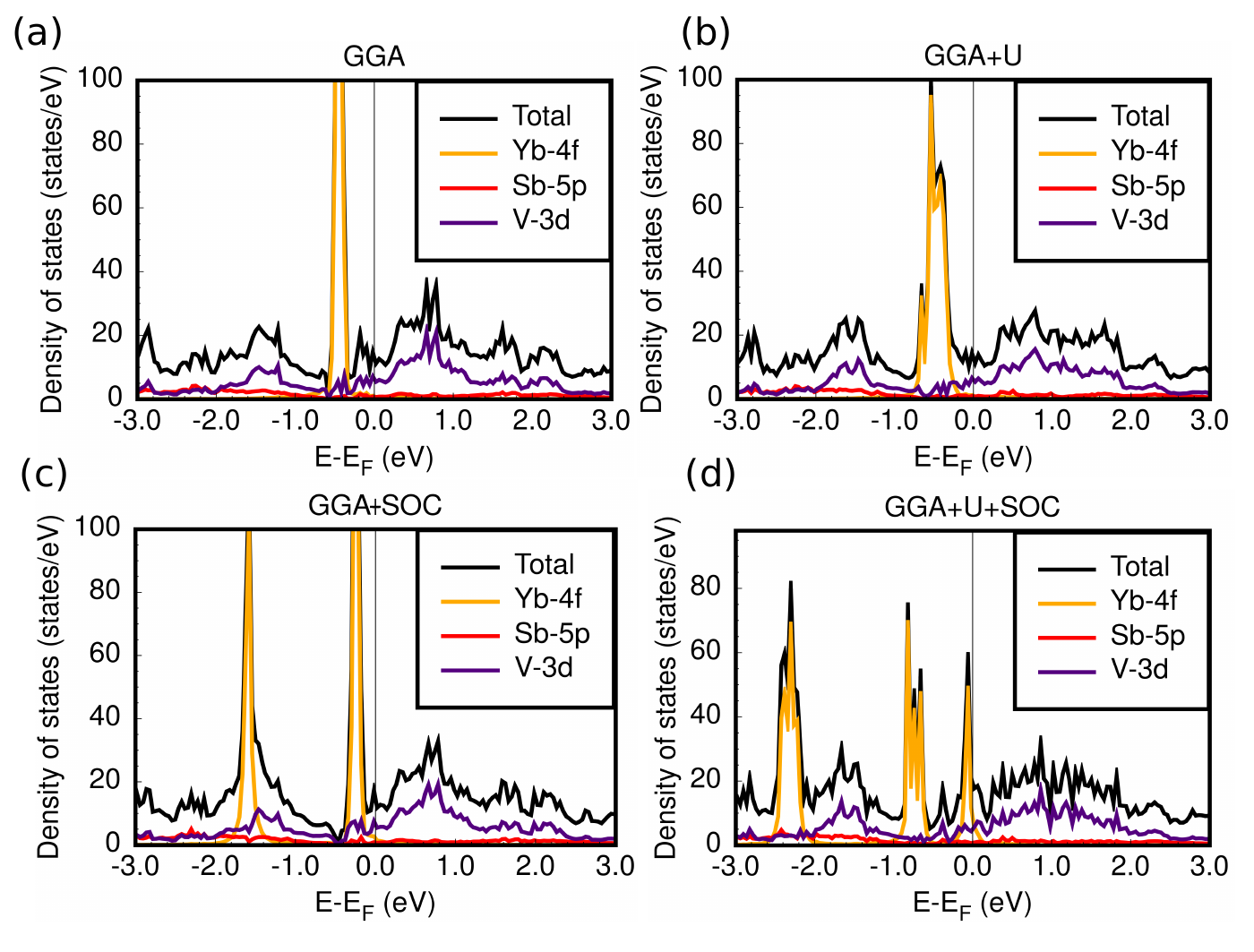} 
		\caption{Comparison of the total and partial Density of States (DOS) as a function of energy (eV) for: (a)  GGA, (b) GGA+U, (c)  GGA+SOC and (d) GGA+U+SOC. The Fermi level is set as a reference energy, indicated with solid lines.} 
		\label{fig:bandcompare}
	\end{figure*}
	
	The unit cell of YbV$_3$Sb$_4$ features zigzag layers of Yb atoms forming a Kagome lattice (depicted in gold), interleaved with V (blue spheres) and Sb (red spheres) atoms arranged in a pattern combining hexagonal motifs and layered structures (see Fig.\ref{fig:four_images}(a)). The Kagome structure is slightly distorted as shown in  Fig.\ref{fig:four_images}(b). The V atoms form a kagome net with V–V bond lengths of $2.81\,\mathrm{\AA}$ and $2.83\,\mathrm{\AA}$, indicating that every bond deviates by at most $0.1\,\mathrm{\AA}$ from the ideal kagome lattice. The zigzag chain running along the $a$-axis forms an angle of $93.90^\circ$, with intrachain atomic separations of approximately 3.9~\AA, as illustrated in Fig.~\ref{fig:four_images}(c),whereas the nearest neighbor interchain spacing is $\sim5.7\,\mathrm{\AA}$.

	\begin{figure*}[htbp]
		\centering
		\includegraphics[width=\textwidth]{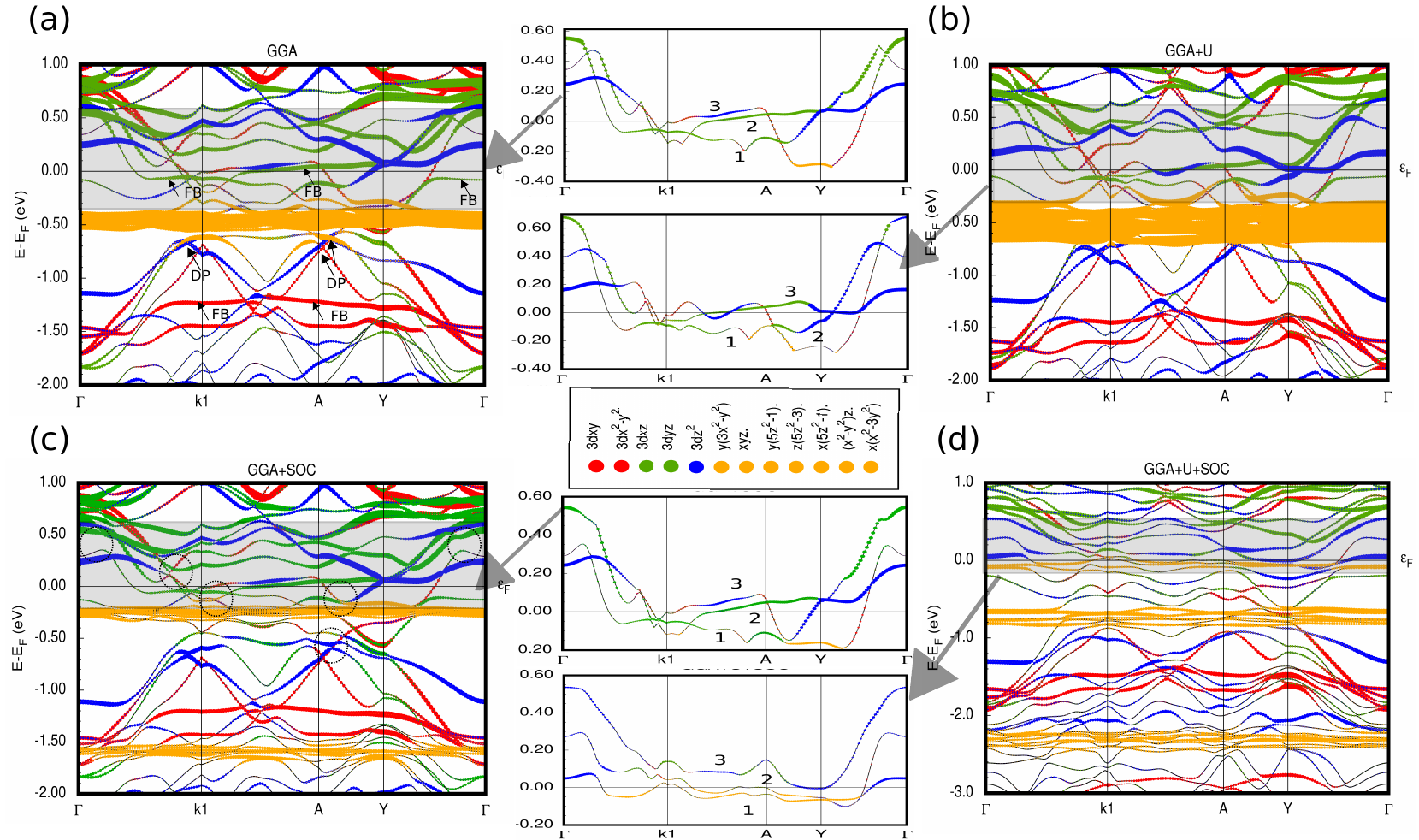}
		\caption{DFT calculated electronic properties of YbV$_3$Sb$_4$, showing orbital resolved band structures obtained by: (a)  GGA, (b) GGA+U, (c) GGA+SOC and (d) GGA+U+SOC. } 
		\label{fig:large_figure}
	\end{figure*}
	
	\subsection{Electronic Properties}
	To perform the comprehensive study we have computed the electronic properties of YbV$_3$Sb$_4$, that includes the band structure and  Density of States (DOS) assuming different approximations GGA, GGA+U, GGA+SOC and GGA+U+SOC within the DFT formalism. We also verified that the ground state is non magnetic, regardless of the treatment of the electronic correlations and spin-orbit coupling, in accordance  with the experimental report.~\cite{ortiz2023ybv}
	
	In a first step we performed the comparative study of electronic properties by calculating the density of states and band structure from GGA and GGA+U. In both  cases YbV$_3$Sb$_4$ is found to be metallic [Fig.\ref{fig:bandcompare}(a,b) and Fig.\ref{fig:large_figure}(a,b)]. The DOS reveals that V-3$d$ states especially $d_{xz}$, $d_{yz}$, $d_{xy}$ and $d_{x^2 - y^2}$ orbitals are predominant at the Fermi energy (E$_F$). Flat-band (denoted by FB) like features are observed near E$_F$, mainly contributed by the $d_{xz}$, $d_{yz}$ (green line) and states below Fermi level at -1.5 eV is contributed by $d_{xy}$ and $d_{x^2 - y^2}$ (red line) [Fig.\ref{fig:bandcompare}(a) and Fig.\ref{fig:large_figure}(a)]. While the $d_{z^2}$ states are mostly seen in the intermediate energies. The Yb-4$f$ orbitals are absent at the Fermi level, indicating negligible 4$f$ hybridization. Moreover, characteristic Dirac‐like crossings, coherent of the vanadium kagome network appear in the vicinity of the high‐symmetry points \(k_1\) and \(A\), while the Yb-4$f$ states are localized below -0.3 eV, indicating limited hybridization with the conduction bands.    
	
	On incorporation of the Hubbard $U$ parameter to treat strong electron-electron correlation effect, we have noticed no significant changes rather than the slight shift of the V‑3$d$ bands towards the lower energy below E$_F$ [see Fig.\ref{fig:bandcompare}(b) and Fig.\ref{fig:large_figure}(b)]. Moreover, the Yb‑4f bands show noticeable broadening in which the DOS peak spans roughly from –0.2 eV to –0.8 eV [Fig.\ref{fig:bandcompare}(b)].
	
	On the other hand the GGA+SOC brought major changes in the band profile by splitting the Yb-4$f$ states into two distinct peaks at approximately –0.2 eV and –1.6 eV attributed to F$_{7/2}$ and F$_{5/2}$, respectively, in accordance to Hund's rule [see Fig.(\ref{fig:bandcompare}(c), \ref{fig:large_figure}(c) shown in yellow]. These bands are seen to remain extremely flat. We have also observed the shift of F$_{7/2}$ states closer to Fermi level. The splitting factor ($\Delta \sim$ F$_{7/2}$ $\rightarrow$ F$_{5/2}$ ) is found to be $\sim$1.4 eV, reflecting the strong atomic spin orbit interaction in Yb $f$-orbitals \cite{gui2020novel,guo2018isolated,knebel2006localization}, whereas the V–3$d$ contribution at $E_F$ remains essentially unchanged. The V‐3$d$ bands continue to dominate at E$_F$ exhibiting the band crossings which may be indicative of SOC induced band inversions and emergent topological behavior,see  Fig.\ref{fig:large_figure}(c). The flat‑band like characteristics drawn by V-3$d$ orbital remain protected under SOC. It seems that the V-$3d$ orbitals are more immune to relativistic effects and drive metallic transport channels.
	
	The comparison between the DOS and band structures from GGA+SOC and GGA+U+SOC are shown in Fig.[\ref{fig:bandcompare}(c,d) and \ref{fig:large_figure}(c,d)]. The introduction of Hubbard $U$ to GGA + SOC has split the F$_{7/2}$ multiplet into two sub peaks by $\Delta$=$\sim$0.7 eV, one shifts downwards to $\sim$-0.8 eV and another shifts upwards at the proximity of the Fermi level. The F$_{5/2}$ shift further downwards at $\sim$-2.2 eV, creating 3 distinct energy states. The splitting of $f$-orbitals into three spectral states reflects the interplay of SOC and Hubbard potential (U) under the influence of crystal field splitting \cite{hewson1997kondo,dudarev1998electron,sk2024role}.
	The Yb-4$f$ states at the Fermi level is found to be dominated almost entirely by the $y(5z^2-1)$ orbital character. In contrast, we have not observed major changes in V3$d$ PDOS in both cases while assessing the sensitivity of the Yb4$f$ orbitals to SOC and U when considered simultaneously. As usual, the out-of-plane $d_{z^2}$ orbital becomes strongly dominant near the Fermi energy, whereas the in-plane $d_{xy}$ and $d_{x^2 - y^2}$ orbitals have less contribution. The out-of-plane $d_{xz}$/$d_{yz}$ (green) bands remain dispersive but with noticeably lower weight and narrowed bandwidth around the Fermi level, driving the out-of-plane $d_{z^2}$ orbital and Yb-4$f$ $y(5z^2-1)$ orbital to dominate the low-energy states. Furthermore, we observe band flattening and also gap formations, especially around $k_1$ and $A$ which may indicate correlation-assisted band inversions in the presence of SOC. These observations motivate the analysis of the Fermi surface topology via de Haas–van Alphen oscillations and a direct calculation of the  $Z_2$ invariant  to clearly demonstrate the existence of non‑trivial topological phases.
	
	\subsection{ Fermi surface and de Haas-van Alphen (dHvA) oscillation }
	
	To further understand the low energy electronic structure, we computed the Fermi surface of YbV$_3$Sb$_4$. The Fermi surface was calculated on a self-adjusting k-mesh to ensure maximum accuracy. The Fermi surface was constructed through iterative mesh refinement using an isosurface method. We used a mesh of k points set on a grid of (48 $\times$ 48 $\times$ 48), achieved through three bisections of an initial grid of (6 $\times$ 6 $\times$ 6). 
	
	\begin{figure}[h!]
		\centering
		\includegraphics[width=\linewidth]{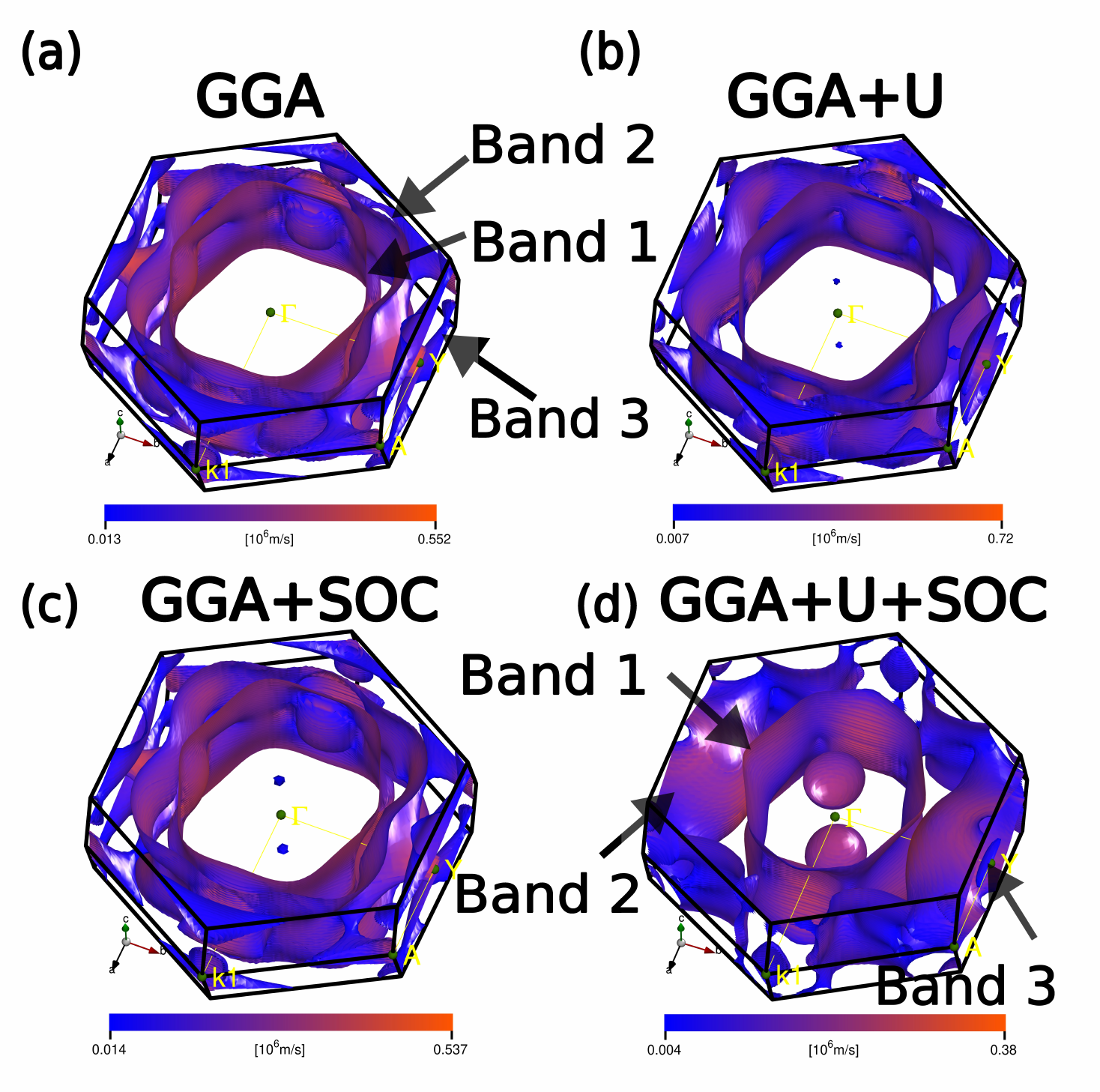} 
		\caption{The Calculated Fermi surface of YbV$_3$Sb$_4$, computed under four approaches: (a) GGA, (b) GGA+U, (c) GGA+SOC, and (d) GGA+U+SOC. Three bands—Band 1, Band 2, and Band 3—cross the Fermi level and contribute to the Fermi surface of this material.The scales at the bottom of each picture are for the Fermi velocity.}
		\label{fig:fermi_surface}
	\end{figure}
	
	{Fig.\ref{fig:fermi_surface} presents the Fermi surface of YbV$_3$Sb$_4$ calculated using the GGA, GGA+U, GGA+SOC, and GGA+U+SOC approaches. The Fermi surface arises from the contributions of three bands: Band 1, Band 2, and Band 3. The detailed band-resolved Fermi surfaces are provided  in Supplementary [see FIG.S2]. Notably, the Fermi surface consists of multiple cylindrical-like sheets centered at the $\Gamma$-point, along with small sheet-like features near the boundaries of the Brillouin zone. The overall topology remains largely unchanged with the inclusion of both U and SOC. However, a small spherical feature at the center becomes more pronounced, and the cylindrical sheets exhibit slight deformation under their combined effect. Similar cylindrical Fermi surfaces have been reported in various titanium- and vanadium-based kagome materials \cite{shrestha2023high,bhandari2024first,bhandari2025pressure,ortiz2021fermi,yang2023observation,wenzel2025interplay,shrestha2023electronic}.}

\begin{figure*}[htbp]
		\centering
		\includegraphics[width=\textwidth]{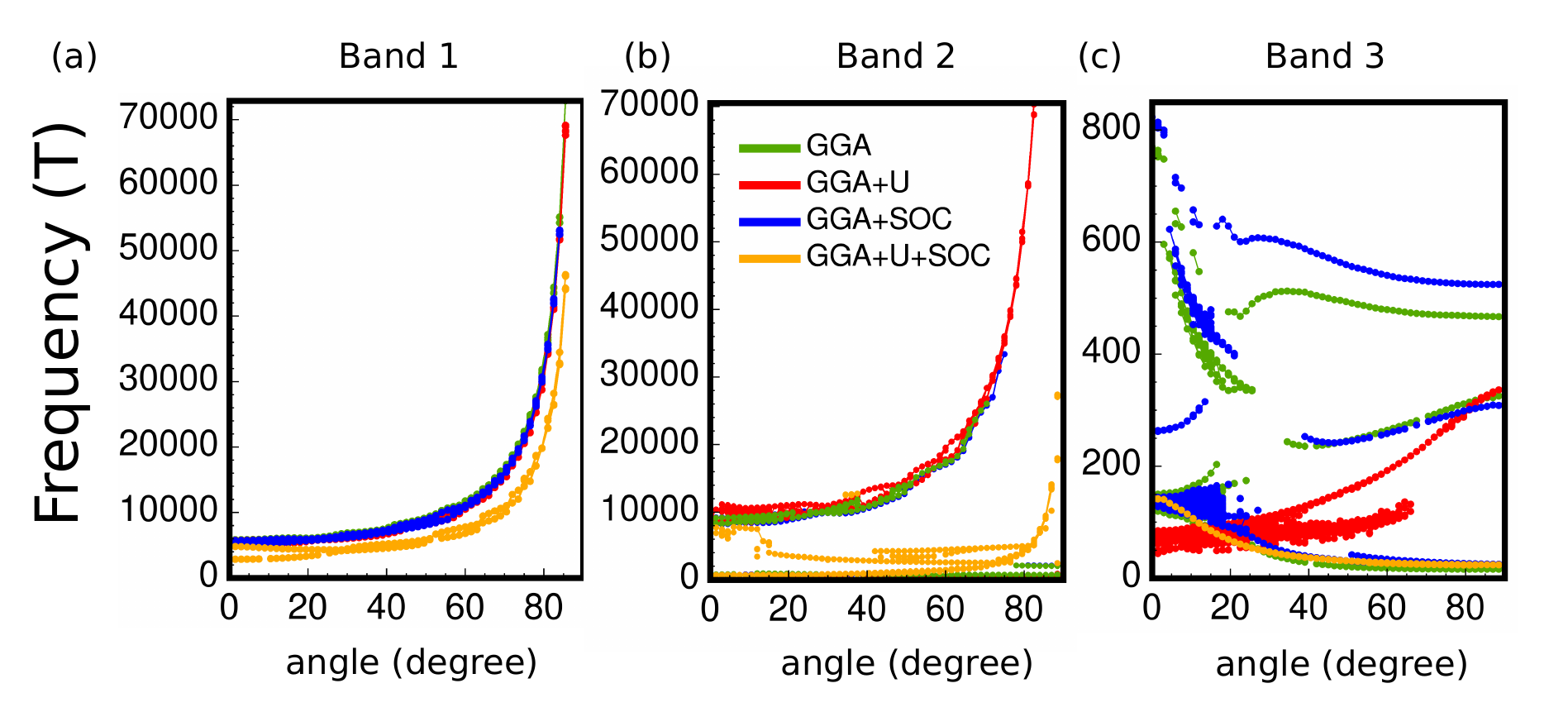} 
		\caption{Theoretical dHvA frequencies of YbV$_3$Sb$_4$ computed using Onsager’s relation at various tilt angles of the magnetic field. Frequencies derived from (a) Band 1, (b) Band 2, and (c) Band 3, incorporating the effects of GGA (green), GGA+U (red), GGA+SOC (blue), and GGA+U+SOC (yellow) in all three panels. The frequencies from Band 1 and Band 2 exhibit nearly parabolic angular dependence, whereas those from Band 3 show a decreasing trend with increasing tilt angle.} 
		\label{fig:medium_figure}
	\end{figure*}

	According to Onsager’s relation Eq. \ref{eq1}, the frequency $F$ of dHvA oscillation is directly proportional to the cross-sectional area \(A\) of the Fermi surface \cite{shoenberg1984magnetic,shrestha2023high,nguyen2022fermiology}:
	\begin{equation}
		F = \frac{\hbar}{2\pi e}A
		\label{eq1}
	\end{equation}
	Therefore, we computed all possible theoretical frequencies by evaluating the cross-sectional areas of the Fermi surface, as shown in Fig.\ref{fig:medium_figure}. As evident in the plot, the frequency increases with increasing tilt angle. Here, the tilt angle is defined as the angle between the c-axis and the direction of the applied magnetic field—0° corresponds to the field aligned along the c-axis, while 90° indicates the field lies within the ab-plane. The frequencies associated with Band 1 and Band 2 increase with tilt angle, reaching values as high as 70 kT. Notably, the angular dependence exhibits a nearly parabolic trend, consistent with expectations for a cylindrical Fermi surface. In contrast, the frequencies derived from Band 3 remain low (below 1000 T) and generally decrease with increasing tilt angle. As expected, except for the GGA+U+SOC case, the frequency curves from different methods largely overlap [see Fig.\ref{fig:medium_figure}(a) and (b)], indicating that the Fermi surface remains largely unaffected by the inclusion of U or SOC individually, but is noticeably modified when both effects are included simultaneously. 
	
	\subsection{$Z_2$ Invariant} 
	
	To investigate the topological nature of YbV$_3$Sb$_4$, four {\it Z$_2$} topological invariants ($ \nu_0 $; $ \nu_1 $$ \nu_2 $$ \nu_3 $) were calculated as proposed by Fu and Kane \cite{kane2005quantum,fu2007topological}. Z$_2$  topological invariants were confirmed with the Wannier charge center (WCC) approach within the PYFPLO code \cite{fploweb}. 
	Importantly, we calculate the Z$_2$ topological invariants for band 2, Fig.\ref{fig:large_figure} under both GGA+SOC and GGA+U+SOC approximations, and in each case found that band 2 which we identify as the highest occupied band exhibits a strong topological index of ($ \nu_0 $; $ \nu_1 $$ \nu_2 $$ \nu_3 $) = (1; 000).  Hence, the compound YbV$_3$Sb$_4$ is categorized to be rich with strong topological behavior providing clear evidence of nontrivial topological band  structures. We further confirmed the automatized calculation of these invariants by visual inspection of the Wannier centres as shown in [Supplementary material Fig.S3]. These Wannier centres on YbV$_3$Sb$_4$  show the evolution of Wannier charge centers as a function of \( k_y \) in different planes and shows a clean straight reference line can be drawn for $\theta$  [See Fig.S3  Supplementary material ], which only crosses this center and hence crosses an odd number of Wannier centers, which results in {\it Z$_2$}= 1. This and the fact that we can visually connect the Wannier center curves in a reasonable smooth way convinces us that the topological indices
	are 1;(000). For trivial cases, there will be a region around $\theta$ where reference line passes without crossing or even number of curves cross in Wannier centre curves. Hence we get 0;(000).
	Based on these results, YbV$_3$Sb$_4$ shows a non-trivial invariant \( Z_2 \) (\( \nu_0 = 1 \)) for the plane \( z_0 \) and trivial invariants for the remaining planes (\( \nu_1, \nu_2, \nu_3 = 0 \)).

	\section{CONCLUSION}
	
	We present the first {\em ab initio} DFT+ Wannier investigation of the vanadium‐based Kagome metal YbV$_3$Sb$_4$,which reveals a nonmagnetic metallic ground state with V-3$d$ derived flat bands and Dirac like crossings near $E_F$ and localized Yb-4$f$ states below it.The inclusion of Hubbard U+SOC splits the Yb-4$f$ multiplets into three distinct peaks without opening a gap, while the V-3$d$ derived band topology remains preserved. The Fermi surface consists of quasi-2D cylindrical sheets which shows slight deformations under U+SOC ,along with the emergence of a small pocket.The  de Haas–van Alphen frequencies computed from these Fermi pockets reach up to $\sim70\,$kT. Importantly, the Wannier‐center analysis reveals a strong $Z_2$ invariant ($\nu_0=1$), establishing YbV$_3$Sb$_4$ as a topological metal. Alltogether, this rare‐earth Kagome compound offers a unique and promising platform for exploring correlated and topological quantum phases.

	\section{ACKNOWLEDGMENTS}
	DPR acknowledges Science \& Engineering Research Board (SERB), New Delhi Govt. of India via File Number: SIR/2022/001150. K.S. acknowledges support from the U.S. Department of Energy, Office of Science, Office of Workforce Development for Teachers and Scientists (WDTS), under the Visiting Faculty Program (VFP) at Los Alamos National Laboratory, administered by the Oak Ridge Institute for Science and Education. Additional support was provided by the Welch Foundation (Grant No. AE-0025) and the National Science Foundation (Award No. 2336011).

	\bibliography{kagomebib}
	
\end{document}